\documentclass[aps,prb,showpacs,twocolumn,showpacs]{revtex4-1}

\usepackage{graphicx}
\usepackage{amsmath}
\usepackage{amssymb}
\usepackage{wasysym}
\usepackage{textcomp}
\usepackage{color}

\definecolor{lila}{rgb}{0.5,0,1}
\newcommand{\bnen}{\begin{equation}}
\newcommand{\eden}{\end{equation}}
\newcommand{\bean}{\begin{eqnarray}}
\newcommand{\eean}{\end{eqnarray}}

\newcommand{\bna}{\begin{array}}
\newcommand{\eda}{\end{array}}

\begin{document}

\title{Characterization of a correlated topological Kondo insulator in one dimension}

\author{I. Hagym\'asi}
\author{\"O. Legeza}

\affiliation{Strongly Correlated Systems "Lend\"ulet" Research Group, Institute for Solid State
Physics and Optics, MTA Wigner Research Centre for Physics, Budapest H-1525 P.O. Box 49, Hungary
}

\date{\today}

\begin{abstract}
We investigate the ground-state of a $p$-wave Kondo-Heisenberg model introduced by Alexandrov and 
Coleman with an Ising-type anisotropy in the Kondo interaction and correlated conduction electrons. 
Our aim is to understand how they affect the stability of the Haldane state obtained in the SU(2) 
symmetric case without the Hubbard interaction. 
By applying the density-matrix
renormalization group algorithm and calculating the entanglement entropy  we show that in the 
anisotropic case a phase
transition occurs and a N\'eel state emerges above a critical value of the Coulomb interaction.
These findings are also corroborated by the examination of the
entanglement spectrum and the spin profile of the system which clarify the structure of each phase. 
\end{abstract}


\maketitle
\section{Introduction} 
Kondo insulators are semiconductors at low temperatures with a few meV gap 
in 
contrast to the conventional semiconductors in which the gap is in the order of a few eV. 
\cite{Kondo_ins:review} It is now generally accepted that strong correlations among the 
electrons lead to the stabilization of this state with such a small gap. \cite{Ueda:review}
\par In the last decade a new group of insulators were discovered, which are gapped in the bulk, but 
they have soft, topologically protected surface states. 
\cite{Moore2010,topological_insulators:review1,topological_insulators:review2}  They are the 
topological insulators whose basic properties can be understood within the framework of ordinary 
band theory. The question naturally arises how the correlation among the electrons affects the 
properties of topological insulators. This aspect has attracted significant attention
\cite{Pesin2010,Maciejko:prl,Ruegg:prl,Tada:prb,Yoshida:prb,Budich:prb,Assaad:jphys,Gregory:prb} 
since the existence of topological phases was suggested in strongly correlated systems. 
\cite{Nagaosa:prl,Chadov2010,Lin2010,Felser:prb,Thomale:prb} These observations made it necessary 
to 
understand, inter alia, the theoretical properties of topological Mott 
\cite{Pesin2010,LeHur2010,Yoshida2012,Yoshida2014,Ning2015} and Kondo insulators.
\cite{Colemanprl2010,Coleman2014,Nikolic:2014,Galitskiprb2014,Alexandrov2015,Galitskiprx2015,
Colemanarxiv2015} The most famous example of topological Kondo insulators is SmB$_6$, which was 
discovered more than four decades ago. \cite{Menth1969} Recent experimental observations 
\cite{Fiskprb2013,Paglioneprx2013,Kim2013} that SmB$_6$ can host robust conducting surface states 
renewed the interest in this compound in particular and in Kondo insulators in general. 
\cite{Vojta2015prl,Sigrist2015prl,Balents2015} The most important aspect of topological Kondo 
insulators is that 
hybridization occurs between $d$ and $f$ bands whose parities are opposite to each other. 
\cite{Colemanarxiv2015} This leads to the appearance of a symmetry-protected topological ground 
state. \cite{Galitskiprx2015}
\par To capture the effects of strong correlations and topology simultaneously, Alexandrov and 
Coleman suggested a Kondo lattice model with a special form of the Kondo exchange. 
\cite{Coleman2014} The model consists of a free one-dimensional electron gas coupled to an $S=1/2$ 
Heisenberg chain via a Kondo exchange with $p$-wave character.  In the conventional 
Kondo lattice model, where the Kondo exchange is on-site, local singlets are formed in the 
strong coupling limit and the boundaries do not play an important role. In contrast, in the present 
case since they become nonlocal, these singlets are broken at the boundary, 
and edge states appear. \cite{Coleman2014} This $p$-wave Kondo-Heisenberg model has been 
investigated with several methods. The large-$N$ expansion has revealed that magnetic end states 
appear at the boundaries. \cite{Coleman2014} Analyzing the ground state with Abelian 
bosonization \cite{Galitskiprx2015} and density-matrix renormalization group (DMRG) algorithm, 
\cite{Gazza2015arxiv} it has turned out that the ground state is actually the Haldane phase. 
\cite{Haldane1,Haldane2} 
It is also interesting to mention that 
Haldane phase has been previously 
observed in the ordinary Kondo lattice model with ferromagnetic Kondo coupling \cite{Ueda1992} and 
in the extended periodic Anderson model with Hund's coupling. \cite{Hagymasi2015}
The facts above clearly indicates that one has to apply sophisticated many-body techniques 
to accurately describe the ground state.
\par In this paper our aim is to study the stability of the Haldane phase against 
perturbations.  The 
most obvious and physically relevant choices are either to introduce anisotropy in the Kondo 
interaction or the inclusion of a Coulomb interaction in the 
conduction band.  The role of the Coulomb interaction was explored in the case of the 
conventional 
Kondo lattice model, \cite{Schork1999,Thalmeier2013} but only the weak coupling limit of the 
present model has been 
studied so far with bosonization. \cite{Galitskiprx2015} In contrast, the 
anisotropy was studied only in the conventional model and its phase diagram was 
determined. \cite{Ueda:anizotrop,Shibata1996363,Novais:anizotrop} We address these
problems using the DMRG algorithm, 
\cite{White:DMRG1,White:DMRG2,schollwock2005,manmana2005,hallberg2006} which is the 
state-of-the-art tool to determine the ground state and it makes possible to go beyond the 
weakly interacting limit. Since the DMRG calculation is closely related to the quantum information 
theory, we can determine the entanglement entropy 
\cite{legeza2003b,vidallatorre03,calabrese04,legeza2006,rissler2006,luigi2008} and entanglement 
spectrum, \cite{Haldane2008} which are very useful to detect quantum phase transitions 
\cite{gu:prl2004,wu:prl2004,yang:pra2005,deng:prb2006} and symmetry-protected topological order. 
\cite{Pollmann2010prb,Pollmann2011prb,Pollmann2012prb} Our analysis of the anisotropic model 
reveals that (i) a phase 
transition occurs as the interaction is increased within the conduction band and the Haldane state 
transforms into a N\'eel state, and (ii) the entanglement spectrum can be used as a fingerprint to 
identify a topological Kondo insulator in one dimension.
\par The setup of the paper is as follows. In Sec. II. the model is introduced and some details of 
the DMRG calculation and the quantum information theory are given. In Sec. III. A our results are 
presented for the isotropic model and we determine the low-lying excitation spectrum. In Sec. III. 
B the role of the anisotropy is addressed using the elements of quantum information theory to 
identify topological order and quantum phase transitions. Finally, in Sec. IV. we give the 
conclusions of this work.
\section{Model} 
The Hamiltonian of the $p$-wave Kondo-Heisenberg model in one dimension can be 
written as:
\begin{equation} 
\label{eq:Hamiltonian}
H=H_c+H_H+H_K,
\end{equation}
where
\begin{equation}
H_c=-t\sum_j\left(c_{j+1\sigma}^{\dagger}c_{j\sigma}+ c_{j\sigma}^{\dagger}c_{j+1\sigma}\right)+U\sum_jn_{j\uparrow}n_{j\downarrow}
\end{equation}
describes the interacting conduction band, with nearest-neighbor hopping $t$ and Hubbard repulsion 
$U$. The Hamiltonian
\begin{equation}
H_H=J_H\sum_j\mathbf{S}_j \cdot \mathbf{S}_{j+1}
\end{equation}
contains the Heisenberg interactions between the $S=1/2$ spins, with $J_H>0$. Finally, $H_K$ couples the two
subsystems via an anisotropic, nonlocal Kondo exchange introduced between electronic and spin 
degrees of freedom, similarly as in Ref. [\onlinecite{Ueda:anizotrop}]:
\begin{equation}
H_K=J_K\sum_{j}\left[\frac{1}{2}\left(S_j^{+}\pi_j^{-}+S_j^{-}\pi_j^{+}\right)+\Delta 
S^z_j\pi^z_j \right],
\end{equation}
where $\Delta$ is the strength of the Ising-type anisotropy, $S^{\pm}_j$ and $\pi^{\pm}_j$ 
($S^{z}_j$ and $\pi^{z}_j$) are the ladder operators ($z$-components) of the spin $\mathbf{S}_j$ and 
the $p$-wave spin density, $\boldsymbol{\pi}_j$:
\begin{equation}
\boldsymbol{\pi}_j=\frac{1}{2}\sum_{\alpha\beta}
p_{j\alpha}^{\dagger}\boldsymbol{\sigma}^{\phantom\dagger}_{\alpha\beta}p_{j\beta}^{
\phantom\dagger } .
\end{equation}
Here $\boldsymbol{\sigma}$ is the vector of Pauli matrices and
\begin{equation}
p_{j\sigma}\equiv (c_{j+1\sigma}-c_{j-1\sigma})/\sqrt{2},
\end{equation}
where $c_{0\sigma}=c_{L+1\sigma}=0$ is assumed.
The present Kondo exchange, in contrast to usual $s$-wave case, couples the localized spins to the 
corresponding $p$-wave spin densities in the fermionic subsystem. As a result, the Kondo interaction 
now contains an exchange term between $\mathbf{S}_j$ and the conduction electron spins 
$\mathbf{s}_{j-1}$, $\mathbf{s}_{j+1}$, where: 
\begin{equation}
\mathbf{s}_j=\frac{1}{2}\sum_{\alpha\beta}c_{j\alpha}^{\dagger}\boldsymbol{\sigma}_{\alpha\beta}c_{j\beta}^{\phantom\dagger}.
\end{equation}
In addition, a new kind of process appears, where an 
electron can hop to its next-nearest neighbor while the intermediate spin is flipped.
\par The 
conventional one-dimensional Kondo lattice model was thoroughly investigated in the 
half-filled 
and in the metallic case \cite{Ueda1992,Ueda1999} and no topological order was found. The role of 
the correlation between the conduction electrons was also addressed and its effect can be 
considered as the renormalization of the spin and charge gaps \cite{Ueda1999}. There are, however, 
only a few studies available on the $p$-wave Kondo-Heisenberg model 
\cite{Coleman2014,Galitskiprx2015,Gazza2015arxiv} and the role of the interaction between the 
conduction electrons has been studied only in the weakly interacting case. Furthermore, the effect 
of the anisotropy in the present model is a completely open question.
\par We applied the DMRG algorithm using the dynamic block-state selection 
approach \cite{DBSS:cikk1,DBSS:cikk2} to study the ground state of the Hamiltonian in Eq. 
(\ref{eq:Hamiltonian}). In our calculation the a priori value of the quantum information loss, 
$\chi$, was set to $\chi=10^{-5}$, and the truncation errors were in the order of magnitude of 
$10^{-7}$. We considered chains with open boundary conditions up to lengths $L=600$.
 In what follows we use the half bandwith $W=2t$ as the energy scale and set $J_H=0.5W$, $J_K=W$ and 
vary the anisotropy and the Hubbard interaction strength. We consider the half-filled case and 
$\Delta\geq 1$.
 \par  It has been shown recently 
\cite{Pollmann2010prb,Pollmann2011prb,Pollmann2012prb} that studying the bipartite entanglement
 of the ground-state wave function is a powerful tool to identify topological phases. Namely, the 
degeneracy of the entanglement spectrum, which is obtained from the Schmidt values, can be used to 
characterize the topological order. Therefore changes in the character of the whole entanglement
spectrum must be accompanied by a phase transition, where either a level-crossing occurs or the 
correlation length diverges. We study the entanglement between the two halves of the chain. To this 
end we  
determine the von Neumann entropy  of the half chain 
(block entropy): \cite{legeza2003b,vidallatorre03,legeza2006}
\begin{gather}
s(L/2)=-{\rm Tr } \ \rho_{L/2}\ln\rho_{L/2}=-\sum_j\Lambda_j\ln\Lambda_j,
\end{gather}
where $\rho_{L/2}$ is the reduced density matrix belonging to the half chain and $\Lambda_j$ are 
the corresponding eigenvalues which are the squares of the Schmidt values. 
\section{Results} 
\subsection{Isotropic case} 
\begin{figure}[t]
\includegraphics[scale=0.6]{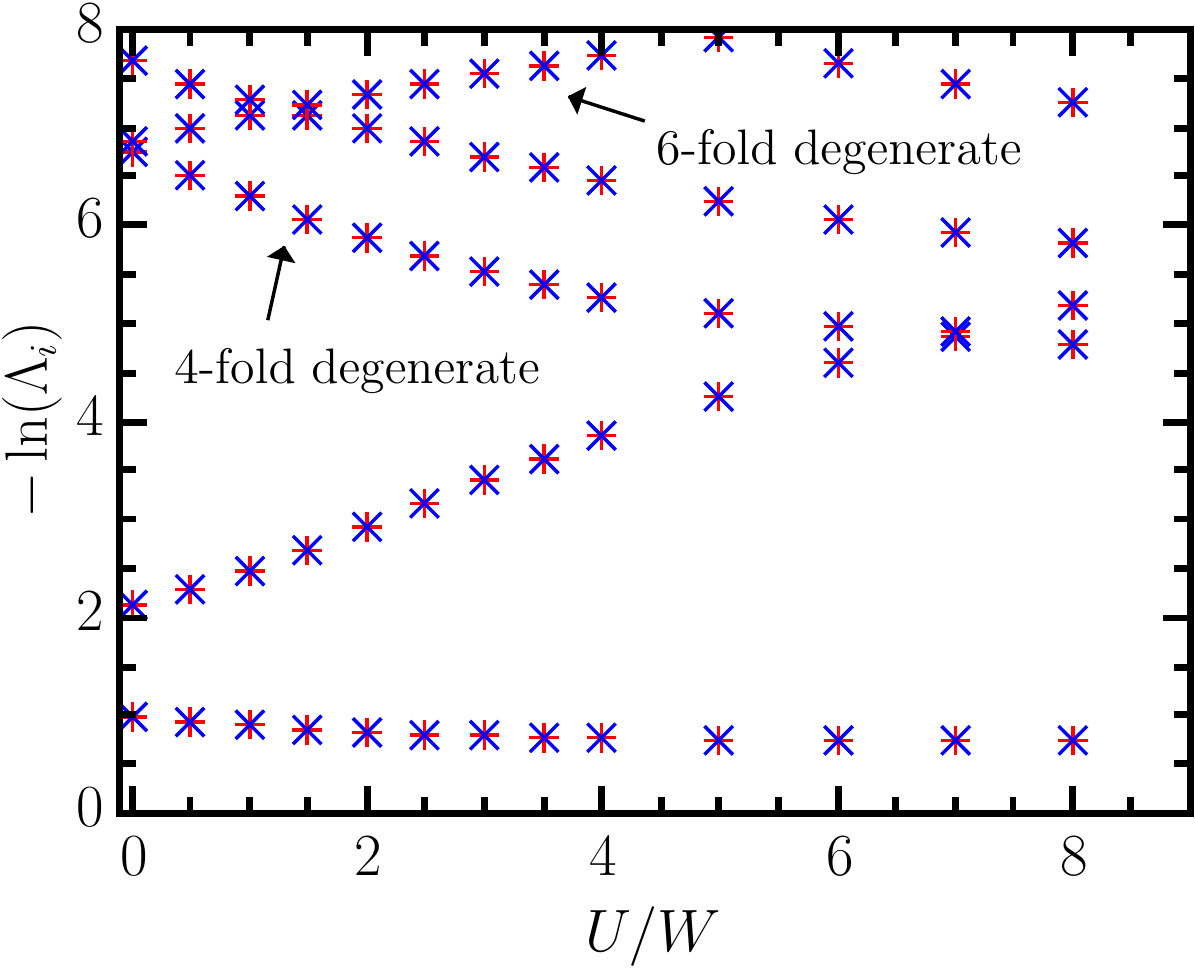}
\caption{(color online) The low-lying entanglement spectrum in the isotropic case extrapolated to 
the thermodynamic limit for $J_K=W$ and $J_H=0.5W$. For better visibility of the 
degeneracies, the subsequent eigenvalues are denoted by $+$ and $\times$ symbols, respectively. In 
certain cases, where the multiplicity is larger than 2, the degree of degeneracy is given 
explicitly in the figure. }
\label{fig:entanglement_spect_iso}
\end{figure}

\begin{figure}[t]
\includegraphics[scale=0.6]{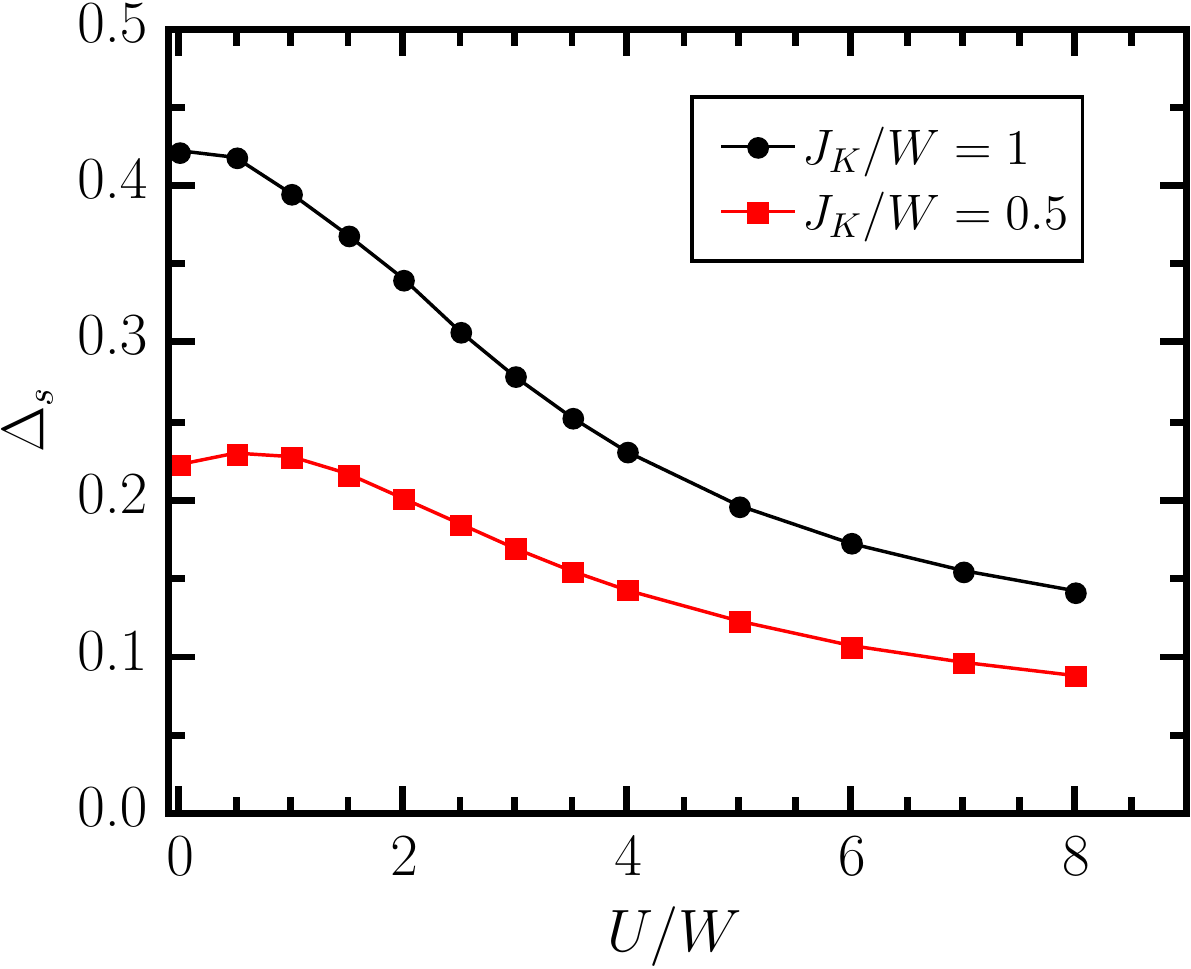}
\caption{(color online) The spin gap (extrapolated to the thermodynamic limit) in the isotropic 
model as a function of $U$ for varous $J_K$ and $J_H=0.5W$. The lines are guides to the eye.}
\label{fig:spin_gap}
\end{figure}

First, we examined the SU(2) symmetric case ($\Delta=1$) and 
calculated the entanglement spectrum as a function of $U$ as it can be seen in 
Fig.~\ref{fig:entanglement_spect_iso}.
It is known that one hallmark of the Haldane phase is its exactly evenly degenerate entanglement 
spectrum, \cite{Pollmann2010prb} which is obvious from Fig.~\ref{fig:entanglement_spect_iso}.
Note that our result provides a further independent evidence for the Haldane phase at $U=0$, which 
was 
previously identified by its non-vanishing string order 
parameter. \cite{Gazza2015arxiv} Besides that it 
is clearly 
observed from Fig.~\ref{fig:entanglement_spect_iso}, that the  Haldane phase is realized not only 
for 
small $U$ but for any $U\geq0$, which is clearly 
beyond the validity of the bosonization approach. \cite{Galitskiprx2015} One can naturally ask how 
the energy scales are modified after switching on $U$. The spin gap, $\Delta_s$, which in our 
case can be 
identified with the Haldane gap (the singlet-quintet gap), decreases as is shown in 
Fig.~\ref{fig:spin_gap}.
This can be understood from the large-$U$ limit of the present model since it becomes equivalent 
to a diagonal spin ladder where the $H_c$ Hamiltonian can be approximated with a $S=1/2$ 
Heisenberg model with $J_c\sim1/U$.
\cite{legeza:ladder}
It is interesting to remark that the Hubbard interaction leads to the increase of the spin 
gap in the conventional model. \cite{Ueda1999}
 \subsection{Anisotropic case} 
\begin{figure}[ht]
\includegraphics[scale=0.6]{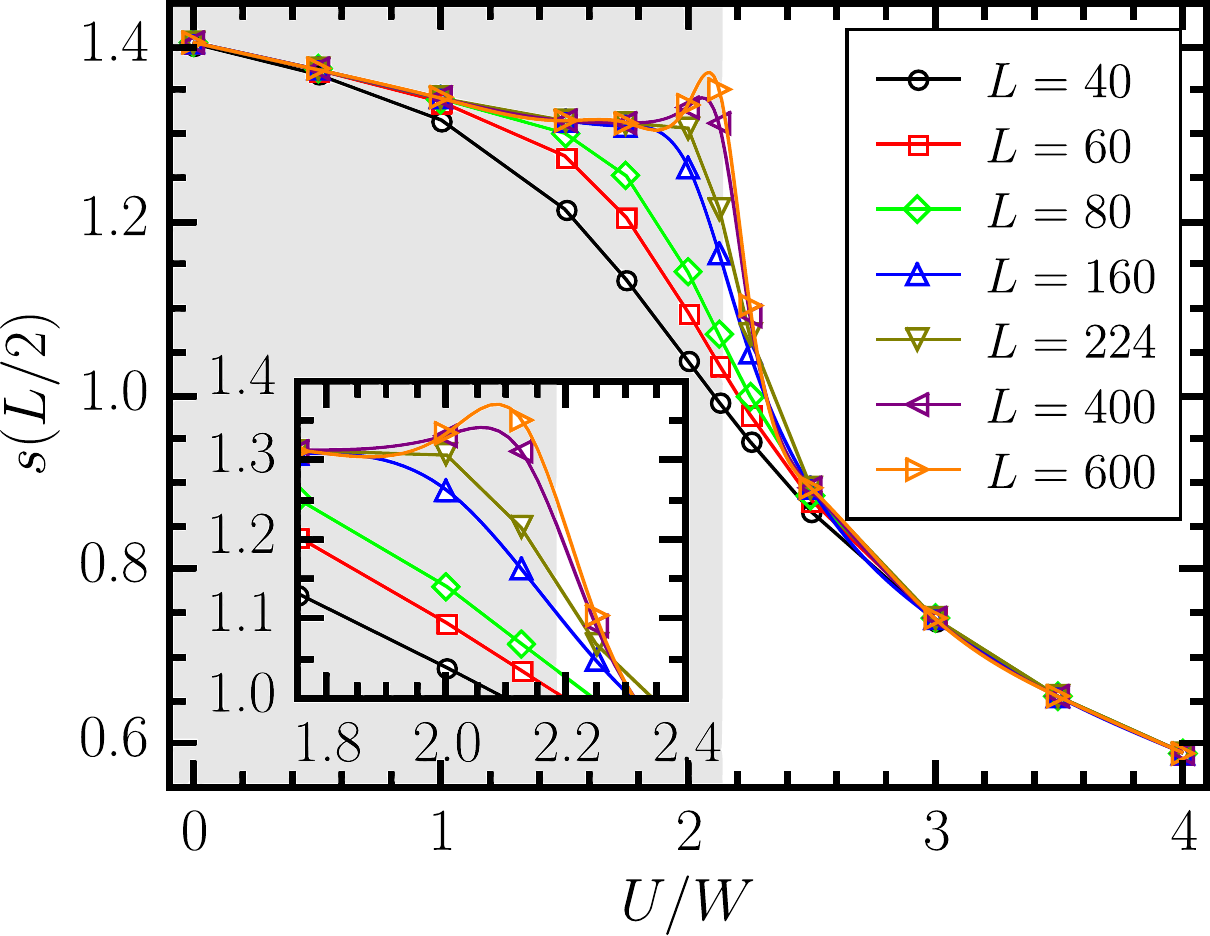}
\caption{(color online) The main figure shows the block entropy in the anisotropic case as a 
function of $U$ for several 
chain lengths and $\Delta=2$, $J_K=W$, $J_H=0.5W$. The lines are the spline fit to the data points. 
The inset 
shows the enlarged region around the critical point.}
\label{fig:block_entropy}
\end{figure}

\begin{figure}[!htb]
\includegraphics[scale=0.6]{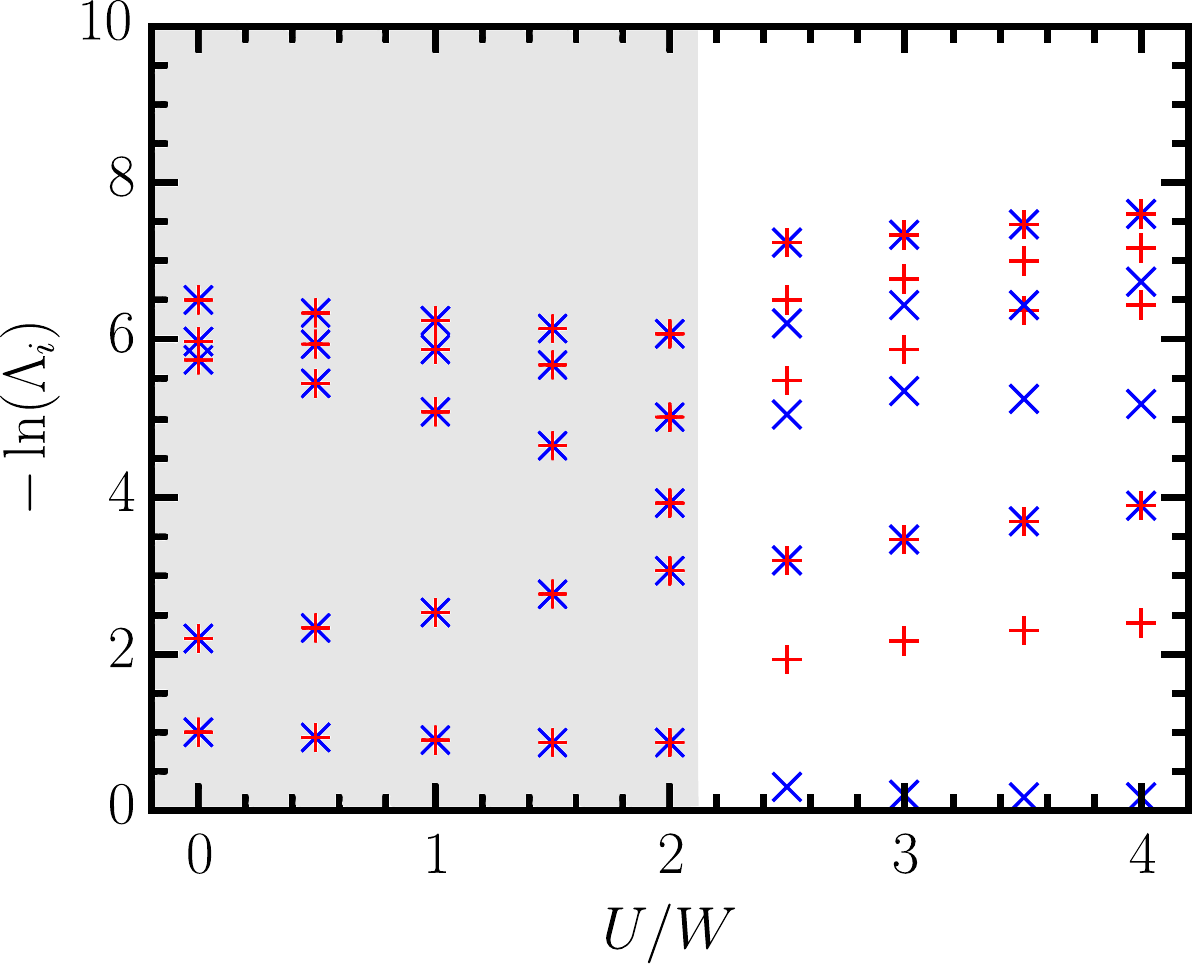}
\caption{(color online) The ten largest values of entanglement spectrum extrapolated to the 
thermodynamic limit as a function of $U$ for $\Delta=2$, $J_K=W$, $J_H=0.5W$. For better visibility 
of the 
degeneracies, the subsequent eigenvalues are denoted by $+$ and $\times$ symbols, respectively.} 
\label{fig:entanglement_spect}
\end{figure}

 As a next step we investigated what happens when the Kondo 
interaction becomes anisotropic ($\Delta\geq1$). 
We calculated the block entropy for several chain lengths as a function of $U$, which is shown in 
Fig.~\ref{fig:block_entropy}. 
It is clearly observed that for short chains they do not exhibit anomalous behavior, however, a 
peak 
is developed as the chain length
is increased. We know that extrema in
the block entropy can be attributed to quantum critical
points \cite{legeza2006} if they evolve into anomalies in the thermodynamic limit. In the present 
case due to the anisotropy, a phase transition occurs at 
$U_{\rm c}\approx2.13W$, which separates two distinct phases. 
To detect symmetry-protected 
topological phases,
we calculated again the entanglement spectrum, and its low-lying part is shown 
in Fig.~\ref{fig:entanglement_spect} as a function of $U$.
It is clear from Fig.~\ref{fig:entanglement_spect} that the Haldane phase at $U=0$ is not destroyed 
by the anistropy, it survives even at $\Delta=2$. Switching on $U$ does not lead to drastic 
changes as long as $U<U_{\rm c}$.
Above this value a new phase emerges, where the 
topological order is no longer present, since both odd and even degeneracies occur in the 
entanglement spectrum. 

\par
\begin{figure}[!htb]
\includegraphics[scale=0.6]{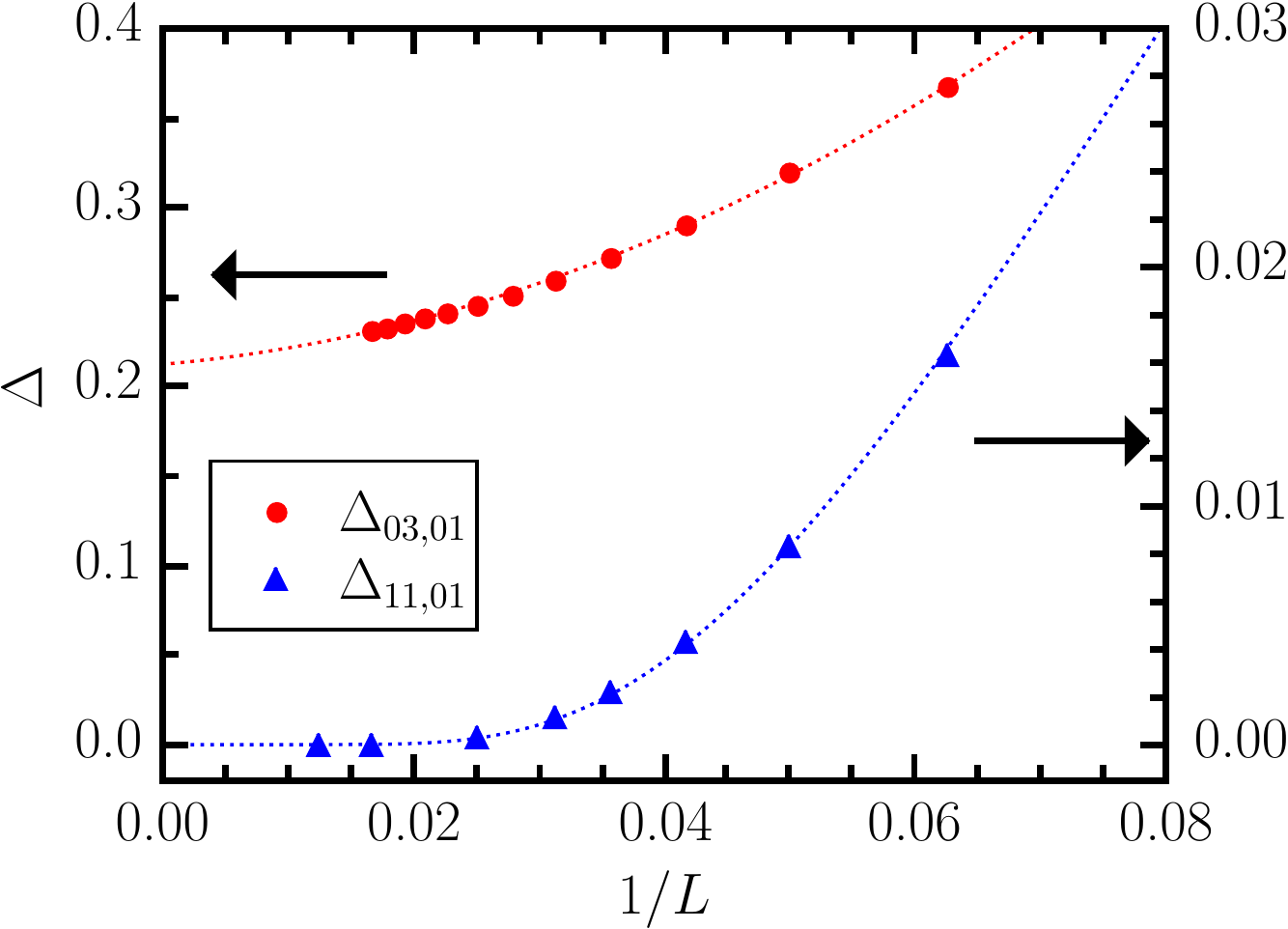}
\caption{(color online) Various gap values as a function of inverse chain length for $U=0$ and 
$J_K=W$, $J_H=0.5W$ for chains with up to $L=80$ lattice sites. 
The dotted lines denote the best fits to the data using Eqs
(\ref{eq:exp_fit}), (\ref{eq:quadratic_fit}) (see text).}
\label{fig:gaps_haldane}
\end{figure}

 To gain further insight into the properties of 
the two 
phases, we have calculated the low-lying excitation spectrum in each phase for $\Delta=2$. More 
precisely, 
we 
considered the following energy gaps:
\begin{equation}
 \Delta_{KL,MN}=E_L(T^z=K)-E_N(T^z=M).
\end{equation}
Here $T^z$ denotes the $z$-component of the total spin:
\begin{equation}
\mathbf{T}=\sum_j\mathbf{T}_j\equiv \sum_j( \mathbf{S}_j+\mathbf{s}_j),
\end{equation}
furthermore, $E_L(T^z=K)$ denotes the $L$th energy level ($L=1,2,\dots$) in the sector
$T^z=K$.
First, we consider the energy spectrum for $U=0$, which is shown in 
Fig.~\ref{fig:gaps_haldane}.
In the Haldane phase the ground state is expected to be fourfold degenerate if open boundary 
condition is 
applied.  
For short chains a finite gap is observed, since the end spins are correlated, however,
it closes exponentially, as the chain length is increased. Therefore these values can be
fitted by:
\begin{equation}
\label{eq:exp_fit}
 \Delta_{11,01}(L)=\Delta_{11,01}+A\exp(-\xi/L),
\end{equation}
where fitting paramers are $\Delta_{11,01}=8(6)\cdot10^{-5}$, $A=0.266(4)$ and $\xi=0.174(1)$.
Since the fitted 
value of $\Delta_{11,01}$ lies within the error margin of the energy values (determined by 
truncation 
error), we conclude that it is zero. Similar considerations apply for $\Delta_{02,01}$ (not shown). 
That is, the fourfold degeneracy is fulfilled in the thermodynamic limit.
On 
the other hand, the gap between the
ground state and the third excited state in the $T^z=0$ sector is finite and its
thermodynamic value can be obtained using the usual quadratic fit: 
\begin{equation} \label{eq:quadratic_fit}
 \Delta_{03,01}(L)=\Delta_{03,01}+B/L+C/L^2,
\end{equation}
with best fit parameters $\Delta_{03,01}=0.212(1)$ $B=0.60(8)$ and $C=30(1)$. This gap can 
be identified as the Haldane gap in the anisotropic system. 
\par
\begin{figure}[!ht]
\includegraphics[scale=0.6]{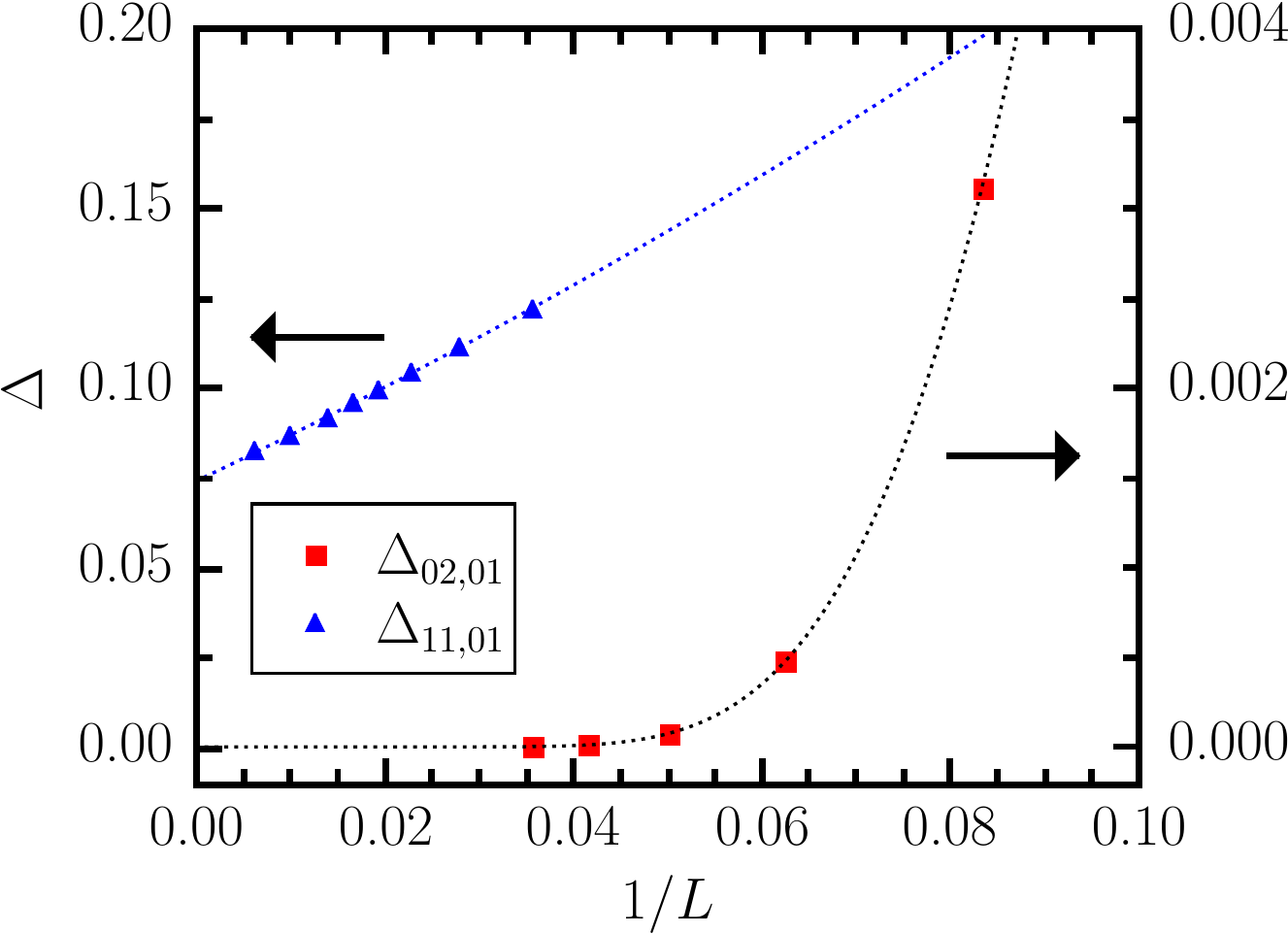}
\caption{(color online) Various gap values as a function of inverse chain length for
$U=3W$ and $J_K=W$, $J_H=0.5W$ {for chains with up to $L=160$ lattice sites}. 
The dotted lines denote the best fits to the data using
Eqs. (\ref{eq:exp_fit}), (\ref{eq:quadratic_fit}) (see text).}
\label{fig:gaps_neel}
\end{figure}

 As a next step we consider what happens when \mbox{$U>U_{\rm c}$}. From 
Fig.~\ref{fig:gaps_neel}, it is seen that the ground state is twofold degenerate and the gap
$\Delta_{11,01}$ remains finite. This was concluded from the size dependence of the gaps 
$\Delta_{02,01}$ and $\Delta_{11,01}$ which can be fitted using Eqs. (\ref{eq:exp_fit}) and 
(\ref{eq:quadratic_fit}), respectively. The best fit parameters in this case are $A=0.861(1)$, 
$\xi=0.4685(1)$ and $\Delta_{02,01}=4(1)\cdot10^{-7}$ with Eq. (\ref{eq:exp_fit}), while 
$\Delta_{11,01}=0.0743(3)$, 
$B=1.24(3)$ and $C=2.7(8)$ values are obtained with Eq. (\ref{eq:quadratic_fit}). The tiny gap 
in the 
former case is considered to be zero based on the previous arguments.
\par
\begin{figure}[t]
\includegraphics[scale=0.6]{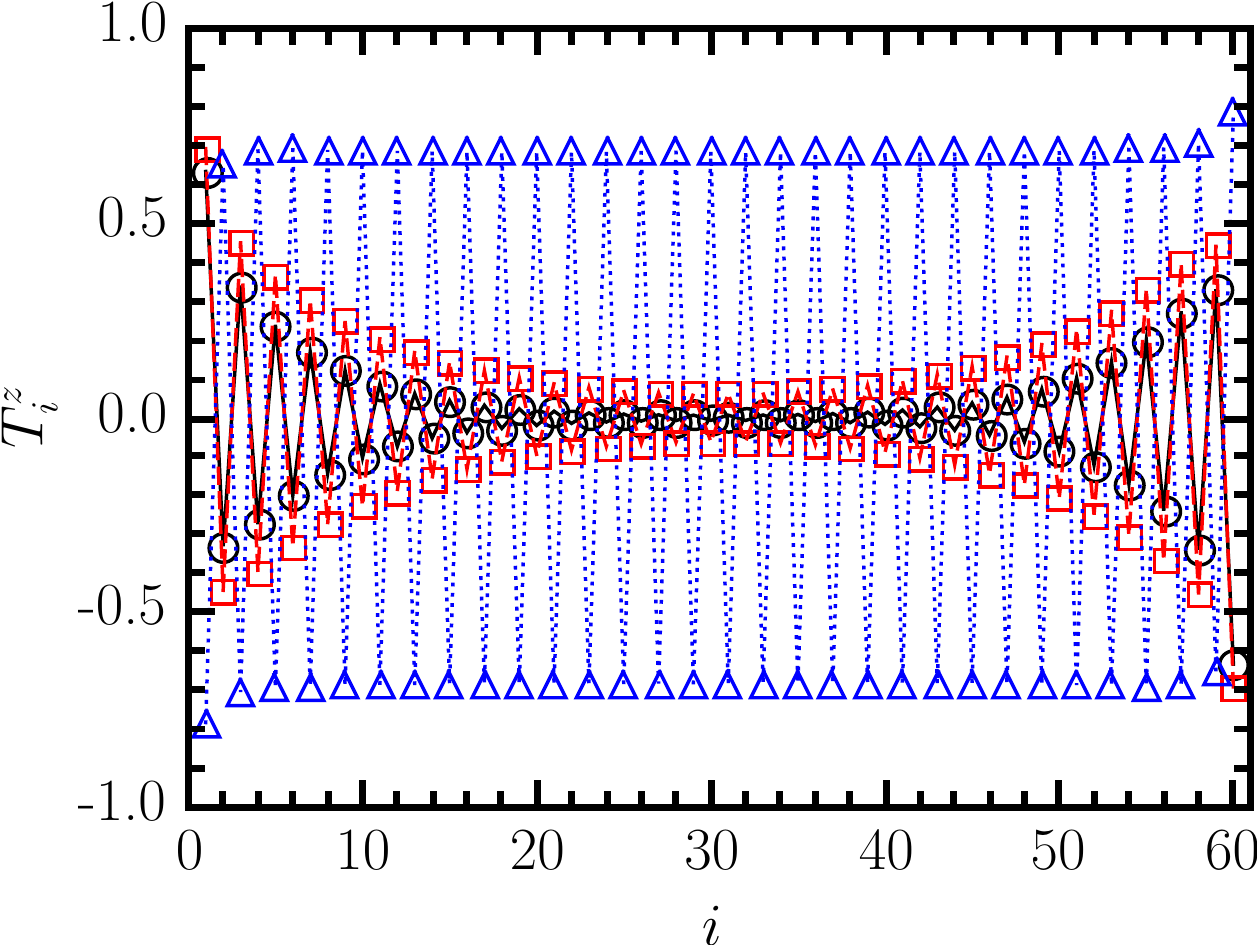}
\caption{(color online) The total spin $z$-component ($T_i^z$) along 
a finite chain with $L=60$ sites and 
for different 
values of $U$. The symbols \textopenbullet, \textcolor{red}{\Square} and 
\textcolor{blue}{$\triangle$} denote $U/W=0$, $1$ and $3$, respectively. The other 
parameters are $J_K=W$, $\Delta=2$, $J_H=0.5W$ in all cases.}
\label{fig:spin_profile}
\end{figure}

\begin{figure}[!ht]
\includegraphics[scale=0.6]{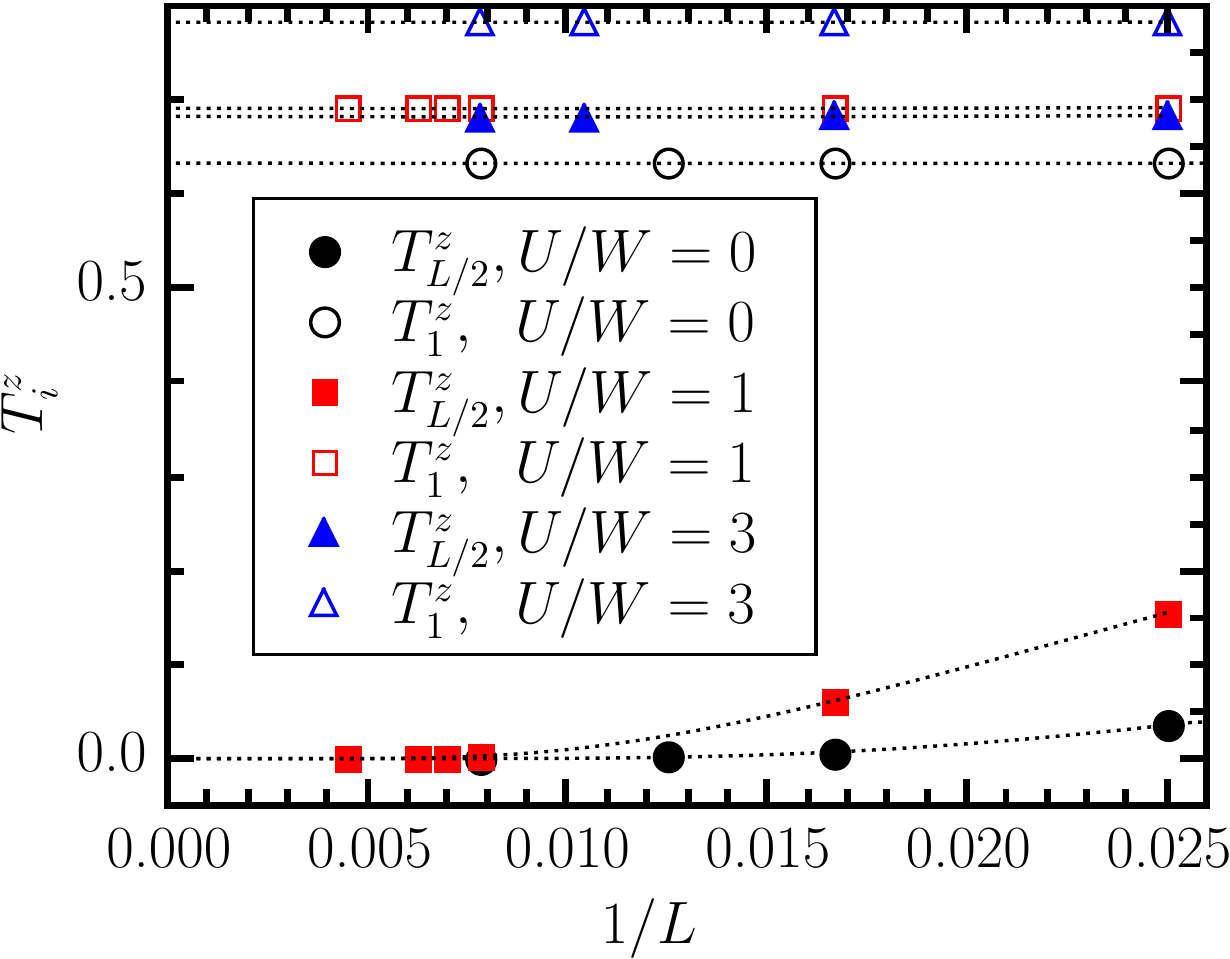}
\caption{(color online)  Finite-size scaling of 
$T_i^z$ values at the end of the chain (open symbols) and in the 
bulk (filled symbols) for different values 
of $U$ up to $L=224$ sites. The other 
parameters are $J_K=W$, $\Delta=2$, $J_H=0.5W$ in all cases.}
\label{fig:spin_profile_scaling}
\end{figure}

 To explore the intrinsic properties of the new phase for $U>U_c$, we investigated the
$z$-component of the local magnetization, $T_j^z$, along the chain, which is shown in 
Fig.~\ref{fig:spin_profile}.
One can easily see that for $U=0$, finite spin polarization appears only at the edges.
This is a well-known feature of the Haldane phase. Since our calculation was performed
with zero total magnetization, the accumulated $1/2$ spins at the boundaries are aligned
opposite to each other. As $U$ is increased the edge spins penetrate more and more into the
bulk. To eliminate the finite-size effects we investigated how the local magnetization
behaves in the thermodynamic limit both in the bulk and at the edges. 
It turns out from 
Fig.~\ref{fig:spin_profile_scaling} that a finite 
spin polarization 
appears in the bulk
for $U>U_{\rm c}$.
That is, the ground state for $U>U_{\rm c}$ becomes a N\'eel state,
which is expected to be twofold degenerate, \cite{Degli:review} as it was confirmed by the 
analysis of the gaps.
The topological Kondo insulator becomes a topologically trivial bulk insulator as the interaction 
is increased.
\par
\begin{figure}[ht]
\includegraphics[scale=0.6]{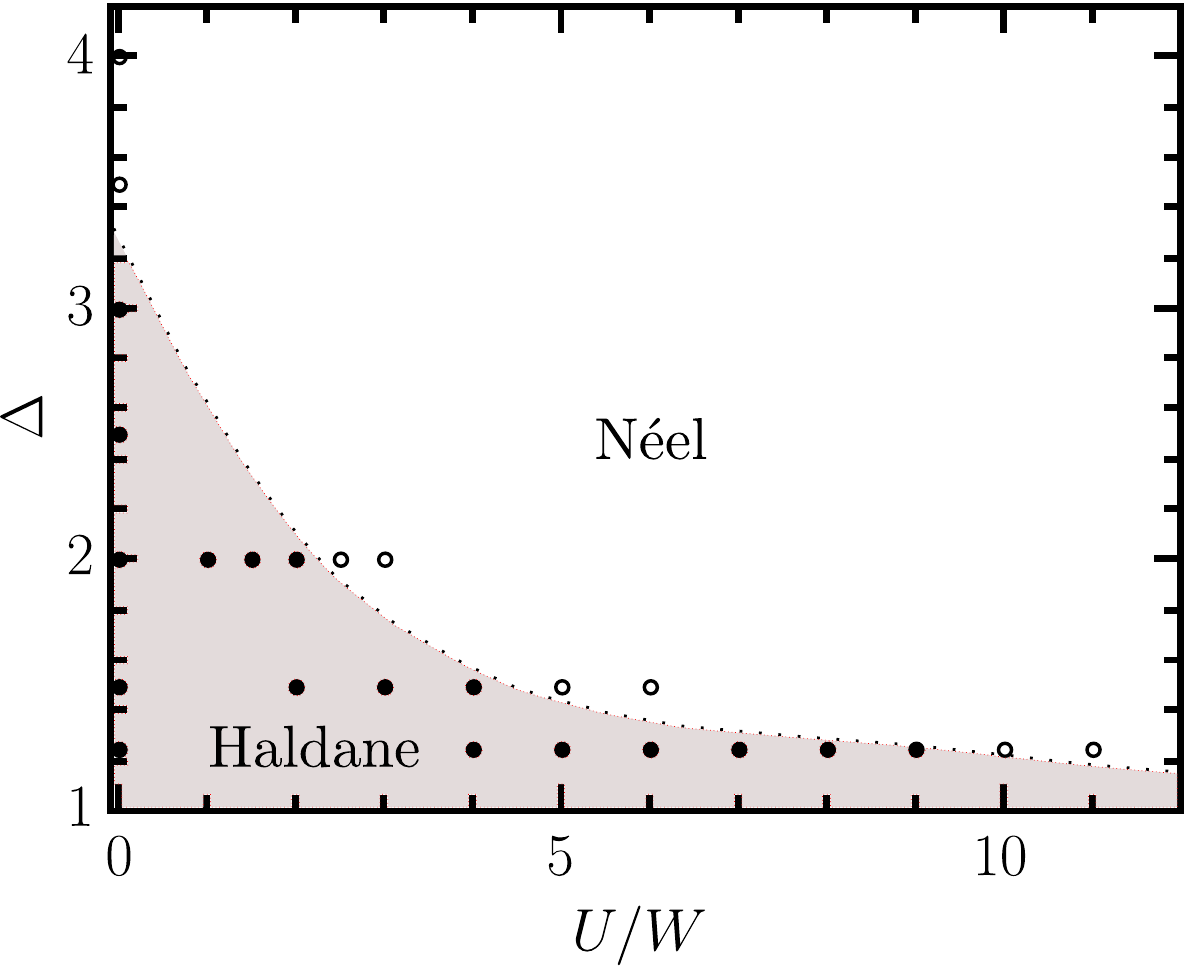}
\caption{The phase diagram of the anisotropic model as functions of the anisotropy and the Hubbard 
interaction strength for  $J_K=W$ and $J_H=0.5W$. The filled and open circles correspond to 
Haldane and N\'eel ground state, respectively. The dotted line denotes the approximate phase 
boundary.}
\label{fig:phase_diagram}
\end{figure}

 Finally, we 
investigated how the critical $U$ depends on the anisotropy. The results are summarized in the 
phase diagram in Fig.~\ref{fig:phase_diagram}.
It is interesting to note that for  a weakly correlated conduction band the Haldane phase extends  
to much larger anisotropy values, than for strong interaction. Note that the phase boundary between 
the Haldane and N\'eel state in the large-$U$ limit remains at $\Delta>1$, similarly as in the case 
of the $S=1$ $XXZ$ chain. \cite{Tasaki:prl}

\section{Conclusions} We investigated the ground state and the first few 
excited states of a one-dimensional  $p$-wave Kondo--Heisenberg model. We demonstrated that the anisotropic Kondo interaction and the correlation between conduction electrons affect the 
topological properties of the ground state. In the isotropic case we pointed out that the Hubbard 
interaction does not alter the topological properties of the ground state, however, it reduces 
the Haldane gap. In contrast, in the anisotropic case analyzing the block entropy, we showed that a 
phase transition occurs as $U$ is 
increased. This was also corroborated by the analysis of the gaps and the change in the structure 
of the entanglement spectrum. The investigation of the local magnetization confirmed that the 
N\'eel phase appears in a certain parameter range. The effect of the anisotropy and the Hubbard 
interaction was summerized in a phase diagram. It is worth emphasizing that when the 
Hubbard interaction is weak the Haldane insulator state is realized in a much wider anisotropy range 
 than in the strongly correlated case.

\acknowledgments{This work was supported in part by the
Hungarian Research Fund (OTKA) through Grant Nos.~K 100908 and NN110360. We acknowledge helpful 
discussions with F. Gebhard and J. S\'olyom. }

\bibliography{tki_refs.bib}

\end{document}